# Wargames as Data: Addressing the Wargamer's Trilemma[1]

Andrew W. Reddie, University of California, Berkeley (areddie@berkeley.edu)
Ruby E. Booth, Sandia National Laboratories
Bethany L. Goldblum, University of California, Berkeley
Kiran Lakkaraju, Sandia National Laboratories
Jason Reinhardt, Sandia National Laboratories

Imagine a conference room in the Pentagon. In it, a senior policymaker is participating in a cyber wargame. As the "Orange" player, she must decide whether to launch cyber-attacks against the military bases of her adversaries. She weighs the actions taken by her fellow players over the past three rounds, including her broken alliance with the "Green" player and the invasion of her territory by the "Purple" player. Will her cyber-attacks demonstrate her resolve or risk disproportionate retribution? This kind of decision is of serious concern to policymakers, scholars, and game designers alike: namely, under what conditions do actions taken in cyberspace have the potential for conflict escalation?

A significant challenge for researchers who study cyber escalation is the lack of data to test different theories about this critical intersection of politics, security, and technology. The paucity of data fuels longstanding debates over the nature of cyberspace and how it compares to other domains of conflict and competition. Some analysts argue that action in cyberspace is likely to exacerbate the risk of escalating conflict.[1] Others are more sanguine.[2] Unfortunately for researchers—though fortunately for everyone else—empirical data about cyber actions and effects during war remains limited.[3] Reliable evidence from conflicts such as Syria and Ukraine are, at best, incomplete. In an attempt to resolve this challenge, researchers have undertaken a variety of efforts to both explore and analyze conflict in the cyber domain by generating synthetic data. Wargaming represents an increasingly important tool for this endeavor.[4]

---

[1] Accepted chapter in *Cyber Wargaming: Practical Research and Education for Security*, Georgetown University Press (forthcoming).

In this chapter, we outline "the wargamer's trilemma" to articulate the trade-offs that wargame designers face. Second, we use this trilemma as a lens to examine the use of wargaming by scholars to study statecraft and conflict in the cyber domain. Here we pay particular attention to the wide variety of analytical wargames and scenarios that already exist in the field. Third, we compare wargames to alternative approaches to generate data, specifically, modeling and simulation on the one hand and survey experiments on the other. In the process, we identify the relative strengths and weaknesses of these different methods—noting where they might usefully augment or complement one another. Based on this comparison, we conclude by highlighting the promise of cyber wargaming to explore and analyze important challenges in national and international security.

**The Wargame Designer's Trilemma**

When creating wargames for the purpose of research or analysis, game designers face a trilemma (i.e., a three way dilemma).[5] Analytical utility, contextual realism, and player engagement are all in tension. The trilemma concept, often found in economics, refers to complex problems with three mutually exclusive solutions or objectives.[6] We employ this term in a broad sense to describe a decision space in which game designers must make tradeoffs between important but at least somewhat incompatible goals.

Our trilemma model illustrates the compromises made when generating synthetic data with wargames, given that there are no perfect solutions.[7] We describe these tradeoffs as a game design moves closer to or further from each node, as outlined in Figure 1. We acknowledge that our model is a simplification of a complex tradespace. However, for the sake of simplicity, we argue that each node and axis is important as well as useful to consider. In the next section, we examine each tradeoff in further detail.

*[Insert: Figure 1: The Wargame Designer's Trilemma]*

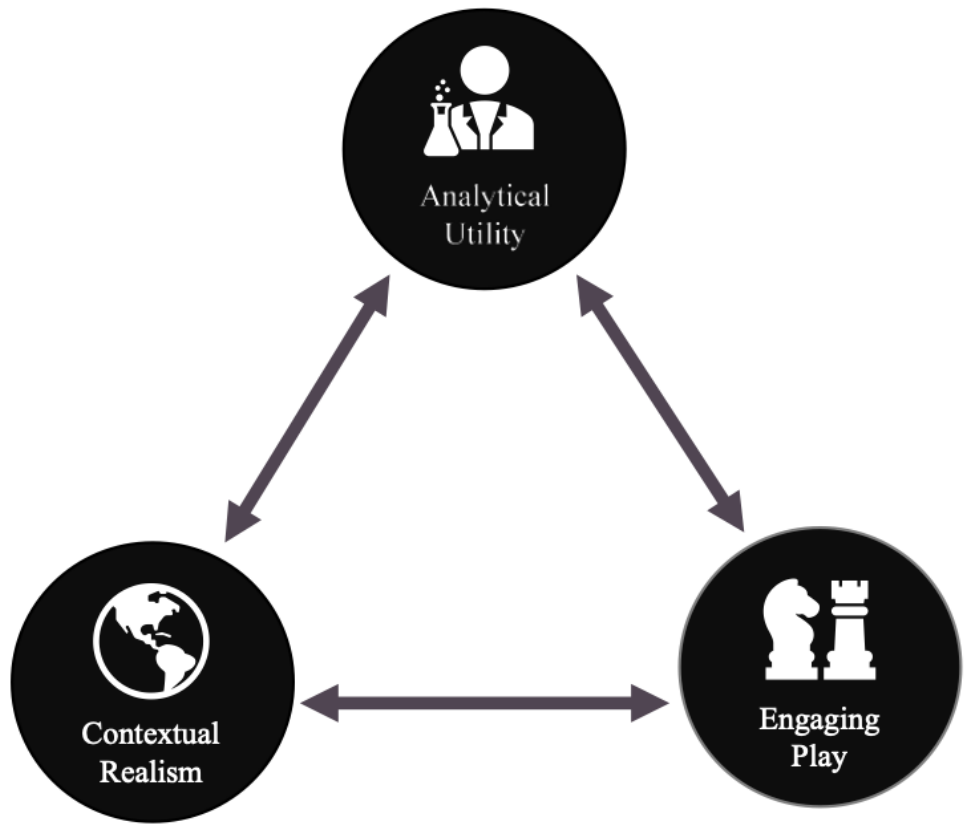

Figure 1: The Wargame Designer's Trilemma

*Balancing Analytical Utility and Contextual Realism*

Wargames, like survey experiments and computer-based models, simplify reality to analyze aspects of it. Yet contextual realism is also essential to a wargame's validity as an analytical tool. Contextual realism refers to the extent that a game adheres faithfully to the real world. On the one hand, wargame designers must retain some fidelity with "the real world" so that players' behavior during gameplay reflects real-world decision making. On the other hand, wargame designers also need to simplify reality; otherwise, it's difficult if not impossible to isolate and test how different causes and effects relate to each other. As a consequence, researchers and wargame designers must omit unnecessary or unwanted elements to remove extraneous artifacts that might confound or confuse answers to their analytical question.

This is an important tradeoff. Increasing fidelity to the real-world, with all its complexity and richness, might confound the analysis. Conversely, by decreasing fidelity through simplification in an effort to generate and capture only the most relevant data, the designer risks losing essential elements that might drive players' behavior in the real world. We are therefore left with the directive: make the game as simple as possible, but no simpler. The difficulty of doing so is one reason why wargaming is sometimes described as an "art" rather than a "science."[8]

Variants of the balancing act between realism and utility are described throughout this book. To strike an appropriate balance, researchers must first identify an appropriate level of fidelity for their analytical objectives; they must design game mechanics that elicit behavior associated with their research question and measure data relevant to the study variables. This is not an easy task. Simplified design may generate clear measurement but lack important contextual details that, when present, alter behavior. In that case, the game may not generate reliable and valid data related to the phenomenon of interest that the researcher believes that it does.

For instance, if a wargame aims to analyze resource management in cyberspace, as is the case in CyberWar 2025, and yet lacks a map, it may be possible to measure actions easily, but players may struggle to grasp the significance of their actions without clear visual cues. However, too much fidelity to real world contexts may cause difficulties of a different type. For example, one of our wargames, SIGNAL, sought to study escalation in a nuclear context.[9] Working on SIGNAL, we found that board design can meaningfully alter player's behavior, as people apply heuristics from the real world or even fictional contexts to determine appropriate strategies (e.g., the large nation at the top of the map is played as "Russia" and the region to its west as "Europe," despite the absence of such labels). Thus, a balance is required. This need for balance between contextual realism and analytical utility is one of the main tensions in effective wargame design.

*Balancing Contextual Realism and Engaging Play*

Adding to the challenge, analytical wargame designers must also consider player engagement, wherein the player is "immersed" in a feeling of involvement and energized focus in game play.[10] As players become immersed, they are able to "suture" themselves into the game world, merging action and awareness within a state of "flow," which increases their concentration on the task at hand.[11] Engaging play is important for analytical and educational wargames alike. Without engagement, players may be desultory or disinterested. They may make random decisions without weighing the ramifications of their actions, resulting in the generation of unrealistic data. In the real world, we must live with the real consequences of our actions, but in wargames, the consequences are relevant in proportion to a player's investment in the game's outcomes.

As with each node pair of the trilemma, to some extent contextual realism and engaging play are opposed. After all, there is minutia associated with every scenario, the boring details of everyday life that don't necessarily add anything to the gaming experience. In a game, by however, these details may be necessary to construct a sufficiently realistic environment. Engaging games abstract away those facets of life that don't matter to the context, while retaining the richness and narrative that gives actions meaning.

Imagine that, in addition to implementation high level policy, our Orange player is forced to choose the rations, uniforms, and bivouac layout for thousands of her troops. Such strict adherence to the concrete details of war would not serve the level of action and reaction—nation versus nation. In fact, that level of commitment to "realism" would render the game unplayable to most audiences—whether driven by a lack of knowledge associated with such fine details or pedantry leading to a lack of engagement. By contrast, one can easily imagine a tactical squad-based game using a few of those same elements to help the player feel connected to "their" troops. Reaching an appropriate balance between realism and engaging play requires building the narrative of the game at a level of abstraction appropriate for the research question.

Here again, some wargamers argue that designing for engaging play is an art that cannot be codified. While there may be some aspects of game design that are truly ineffable, we suggest that it is possible to identify characteristics that tend to lead to player engagement. Games with a compelling "skin," or narrative, are generally more enticing than those in which the skeleton of their mechanics is visible. In wargaming, an attractive board, a well-drawn map, small narrative touches, and a coherent theme all contribute to immersion, or suture. By contrast, an overly simplistic or amateur design can prevent players from ever engaging or pull them out of the game, contributing to careless or unconsidered play due to limited investment in game outcomes.

This is particularly problematic given that an advantage of wargames as an analytical tool compared to alternative data generation processes—discussed below—is their potential to elicit "authentic" behavior from participants. Wargames rely on emotional investment in game outcomes to elicit realistic decision-making during play.[12] Achieving a certain threshold of investment—whether via win conditions, consequential play, or strategic interaction, as well as the game venue and identity of players—is necessary for play that generate useful insights into the questions at hand.

*Balancing Engaging Play and Analytical Utility*

The final tradeoff is between engaging play and analytical utility. From the perspective of analytical utility, the best game is often the most parsimonious (e.g., a simple game with as few decision points as possible that only capture the behavior of interest). However, games that are too simple will not be sufficiently engaging. As with the educational cyber wargaming discussed in Chapter 8, too much analytical utility can detract from player engagement.

All things being equal, the more options for action that are available in a wargame, the larger the sample required to achieve statistically significant results. Adding to this difficulty, more prompts, scenarios, and even cosmetic alterations may result in the need for more players and playthroughs for researchers to assess the robustness and validity of the game as a scientific tool. Too broad an option space can confound analytical utility. Too narrow a space can result in a game that no one cares enough about to play. While there are methods to increase player participation through "crowdsourcing" (through sites such as Amazon Mechanical Turk or university laboratories), engagement is still necessary to give validity to the data generated by their play.

The complexity of game mechanics, such as cards and dice, can also affect engaging play. As is true of player's emotional investment, mechanical complexity has an optimal zone. However, there is no "ideal" amount of complexity or parsimony. The appropriate level of parsimony depends on the game objective, as well as the players and their knowledge and experience in the real-world. Hobbyists who have played many games for entertainment may find more complex rules more engaging than policymakers, for example, who may recoil. Thus, it's important to know who will play the game and why—and, of course, to also play test or beta test the game in advance.

As is true with wargaming in general, so too with cyber wargaming. The trade-offs highlighted by this trilemma are no less significant when wargames are used to generate new data and knowledge about cyberspace. In the following section, we examine how a few cyber wargames address the wargamer's trilemma.

**Wargaming the Cyber Domain**

Let's return to our Orange player in the Pentagon. She finds herself in her strategic conundrum based on the scenario that she and her fellow players have been provided—made real in part by the map and resources that each player enjoys. This begs the question of how a different map or different distribution of capabilities might impact her behavior. These "laboratory effects" driven by design considerations could be significant.[13] For example, are the outcomes of games in which players pick from prescribed options—or menu—for cyber actions analogous to games that allow open response? Similarly, what elements are salient when one seeks to map and model a "cyber" environment?

To engage with these questions, we use the trilemma outlined above to briefly consider three analytical cyber wargames on escalation dynamics. We consider how the respective games create the conditions for meaningful play—paying special attention to cyber aspects of the game.[14]

*The International Crisis War Game*

As described in Chapter 3, the International Crisis War Game generates synthetic data to examine how cyberattacks on nuclear command, control, and communication may affect the probability of nuclear war.[15] In this analytical wargame, the researchers assign players specific roles in a national security cabinet. Working together in these roles as a team, the players respond to a crisis scenario involving an adversary and a third party. The researchers manipulate specific aspects of the scenario as an experimental treatment and then measure how the players' teams respond with regards to nuclear weapons. The game is played over three hours by elite players who possess nuclear and cyber expertise.

The International Crisis War Game is designed to answer a specific research question, expressly designed with analytical utility in mind.[16] As per our trilemma, this choice in turn has ramifications for contextual realism and engaging play. In terms of contextual realism, the game does not involve "real world" countries or geographies. This choice is understandable given the proclivity of players to caricature scenarios when faced with environments that approximate the real world. Despite the fictional countries and map, however, players are provided with detailed information about the countries involved. This information attempts to provide enough realism to support valid analysis.

The game is also one-sided. Rather than allow for strategic interaction, in which players engage with a live opponent, adversary behavior is pre-scripted or pre-determined. Consistent and repeated adversary behavior simplifies data creation and collection, which serves analytical utility. However, the consequences of players' actions are muted or lost in such one-sided play, departing from the reality in which crises are an interactive—and potentially violent—contest of human capabilities and will. This tradeoff risk stands to limit players' engagement. The plot or narrative arc of the game is static. As such, player decisions cannot change the course of gameplay in any meaningful way. Their action in Round 1 of the game does not change the scenario in Round 2. Therefore, if we are to label the International Crisis Wargame associated with trilemma discussed above, it maximizes analytical utility.

*Island Intercept*

In partial contrast, consider the cyber wargame called Island Intercept, which is detailed in Chapter 4. This game explores how cyber capabilities impact the likelihood of state aggression.[17] It centers on a South China Sea scenario, beginning with a Taiwan-centered vignette. Players represent the United States and China. Their actions are adjudicated by an umpire, who uses dice roles to move the game forward.

Unlike the International Crisis Wargame, with its prescribed scenarios, the players decide who "wins" Island Intercept. Each side has cards that describe their specific cyber capabilities (e.g., phishing to gain access to adversary network), as well as their associated costs and benefits (e.g., "political permission level", "likelihood-of-success measure,"). Players decide how to use these capabilities against each other as they work toward the win conditions.

Island Intercept is built with analytical questions in mind: when played repeatedly, it serves as an elicitation tool for a subsequent survey experiment.[18] Yet this cyber wargame hews closely to contextual realism by using real states involved in an-all-to-realistic conflict scenario. It also allows for interaction between the players, so they must deal with the strategic consequences of their respective actions from round to round—an important aspect of engaging play. Moreover, players also select from a wide but not infinite range of possible cyber actions, reducing the complexity of cyber conflict and making the wargame more approachable if not engaging.

These choices about research design might diminish the analytical utility of the game, however. For example, players may to caricature the behavior of Washington and Beijing based on their prior knowledge and assumptions, rather than engage with the strategic dilemmas posed by the game itself. In addition, the characteristics associated with the cyber capability cards may influence players' behavior in ways that might not be as externally valid as hoped. Combining this wargame with a survey experiment goes a long way towards addresses these tradeoffs. Nevertheless, Island Intercept tends to maximize engaged play.

*Cyber Escalation Game*

Finally, consider the seminar-style wargame that Benjamin Jensen and Brandon Valeriano designed to examine cyber escalation.[19] Similar to the International Crisis Wargame, but unlike Island Intercept, players in this game represent a fictional state in conflict with a non-player, peer adversary in a one-scenario, one-sided game. The players receive one of four different briefing packets that represent the experimental treatment. In these packets, Jensen and Valeriano vary whether the player faces a cyber triggering event and whether they have a cyber response option. Data collection seeks to measure whether or not each player responds in an escalatory manner. This game has been played by over two hundred and fifty participants that included "graduate and undergraduate students, government officials, military officers, and private sector employees."[20]

Jensen and Valeriano use briefing packets to approximate the documents used in real life by the U.S. National Security Council. This choice privileges contextual realism, trading off against engaged play. It isn't entirely realistic, however, given the lack of strategic interaction between the players. As a result, this cyber escalation game also tends more towards analytical utility than Island Intercept.

*[Insert: Figure 2: Positioning Exemplars on the Trilemma's Plane]*

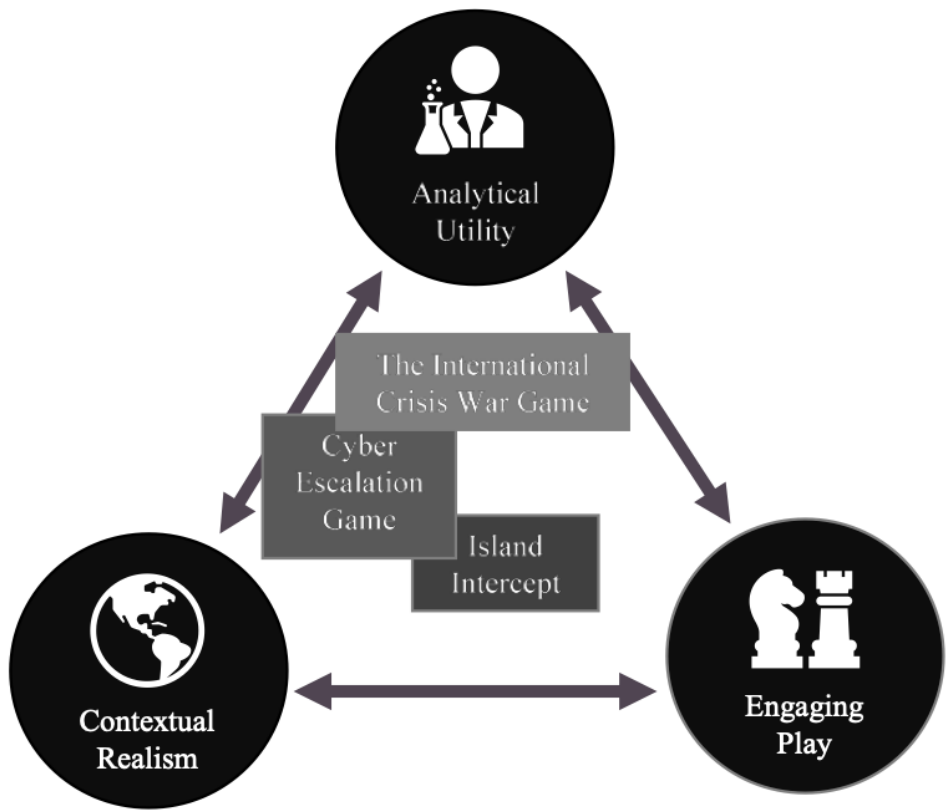

Figure 2: Positioning Exemplars on the Trilemma's Plane

In each of these cyber wargames, we see the wargamer's trilemma at work. Each game sits at a different point in the trilemma simplex, as illustrated by [Insert: Figure 2. This variation has implications for the data generated by each game. This diversity highlights the need to further interrogate how data generation through cyber wargaming relates to other research methods.

**Analyzing the Cyber Domain Across Research Methods**

Cyber wargames offer an exciting tool for observational and, increasingly, experimental work.[21] So, how does our understanding of the Orange player's decision making in a wargame compare to other research methods that analysts use to study cyber competition and conflict? Our trilemma provides a helpful framework for comparing and contrasting observational data, modelling and simulation, survey experiments, and wargames. It also helps us identify potential strengths and weaknesses associated with each approach to data generation.

*Empirical Data*

Scholars typically look to empirical data—whether case-based or statistical in nature—to answer research questions. Questions about cyberspace are no different, although this domain does present some empirical challenges. Much of the theoretical work on cyber operations leans heavily on a limited number of cases to argue that the effects of cyber conflict are unique—or not.[22] Extrapolating from these cases, scholars make myriad claims, including suggestions such as cyberattacks represent a new weapon of mass destruction, that they transform geopolitics due to their relatively low barriers for entry, and that they may offer unique pathways and new motives for conflict escalation.[23]

From our perspective, the challenge with these observational data is two-fold. First, wars that involve the overt use of cyber capabilities have been rare, and "cyber war"—where cyberattacks have physical effects—even more so. Said another way, the "players" in a real war are assumed to be fully engaged, due to the existential consequences, but they rarely play their cyber cards in the open. Second, by its nature, observational data are not created with analysis in mind. Indeed, empirical data represent the most extreme example of the tradeoff between contextual realism and analytical utility, as they contain all the noise and extraneous detail of real life. When researchers convert real events into a case study or data set, they must make choices, and these choices reflect not only their own biases, but also external biases related to how data are recorded, stored, and presented for public consumption.

There are also selection effects associated with empirical data. Existing efforts to collect instances of cyber-attacks reflect information that is publicly available, which may not provide a full picture given the possibility of covert and clandestine operations. For example, cyberattacks by non-state actors who use readily available tools may be overrepresented. These attacks and exploits are less likely to be classified whereas sophisticated state-sponsored attacks are more likely to be kept secret. These selection effects are a concern for analysts seeking generalizable insights. Viewed from the perspective of our trilemma, observational data maximize contextual realism by reflecting events that have occurred, as well as engaging "play" given the high stakes of real life. But they often suffer in terms of analytical utility.

Given these challenges, scholars of cyber conflict have turned to synthetic data generating processes, which may provide more useful data in terms of isolating the impact of various stimuli. Three popular methods for generating synthetic data include modeling and simulation, survey experiments, and wargaming.

*Modeling and Simulation*

When applied to cyber, models and simulations range from focusing on a single defender and attacker to scaled up multi-agents engaging in cyber war.[24] Modeling and simulation tools can be roughly divided into two related but distinct camps—mathematical models of attacker/defender behavior using game theory or more cognitive models and simulations of attacker/defender interactions (e.g., agent-based modeling).[25]

Paring strategic problems down to a relatively parsimonious model represents one major strength of these approaches.[26] Put another way, these methods tend to maximize analytical utility. One challenge, however, is that the assumptions are often baked into these models (e.g., rationality or perfect information). These assumptions may or may not be externally valid. With such assumptions in place, the models may generate perfect, but perfectly inhuman results. Simulations attempt to address this shortcoming by adding variables that can be sampled to provide probability distributions, but they can still miss important contextual factors.

There is also a secondary concern regarding the source material used to create the mathematical models in the first place.[27] For example, several modeling approaches rely on inputs from a limited pool of subject matter experts (SMEs). The resulting models can reflect the proclivities and biases of the SMEs rather than general truths—once again limiting contextual realism and impacting the analytical utility of the results. This can partly be mitigated by "learning" from existing data sets, though selection effects still represent an important concern.[28]

Usually, there is no "player" engagement in these models and simulations of strategic interaction since no humans are involved in generating the data beyond the model or simulation designer. Given the types of strategic behavior that scholars and analysts are nominally interested in within the cyber domain, models that fail to include human behavior are, in our view, problematic. This has led to efforts to bring humans back into the data generating process.

*Survey Experiments*

Survey experiments represent a popular method for including human factors when examining the challenges posed by the cyber domain (and social science in general). They tend to focus on providing only the contextual realism necessary to observe the relevant behavior. Surveys typically pay little attention to how engaging the scenario and questionnaire is for respondents.

In survey experiments, researchers randomly assign participants to a treatment or control group. While experimental treatments vary, they are often expressed by either manipulating the scenario—vignette—or by manipulating the response options provided to the participants.

For example, in a vignette-based study, we might vary the military capabilities that players have among their response options. We might provide some players with offensive cyber capabilities and others with only conventional weapons, for instance. Here, we would be concerned with measuring the different choices selected by respondents that have cyber capabilities against those that do not. Unlike the modeling and simulation methods, survey experiments channel the questions posed by the vignette through human decision-making instead of relying upon a mathematical formula to adjudicate the outcome. The ability to engage with human responses at scale represents a major strength of this method.

A number of scholars have used survey experiments to examine cyber conflict. Some have found that some people support the use of cyber weapons, all else being equal.[29] Others have found that their respondents are reticent regarding the use of cyber capabilities.[30] Yet others have found contingent support for cyber-attacks among the general public—arguing that it depends on attribution, magnitude, or risk perception.[31]

When viewed in terms of the trilemma, surveys designers must address three major methodological concerns. The first concern relates to sampling—an integral aspect of analytical utility. Some scholars suggest that making inferences about strategic decision-making requires engaging with policy elites.[32] Others are much more sanguine about the differences of elites and non-elites.[33] While we do not adjudicate this argument here, further research is likely warranted focused on the sampling effects associated with cyber conflict studies.

The second concern relates to engagement. Survey experiments tend to be short. Tis limited duration and tend to limit the amount of information that surveys can communicate to respondents about the phenomenon under study. Moreover, respondents face minimal consequences for their responses. This limits the engagement or immersion of the survey "experience" and may allow respondents to answer a researcher's questions without weighing the costs and benefits of their actions. Further, evidence suggests that participants tend to respond to survey questions in a manner that reflects their ideals rather than their true behaviors.[34] This bias can be minimized with careful survey design but cannot reliably be eliminated.

Third, survey experiments struggle to incorporate strategic interaction. Many research questions of interest in the cyber domain involve dynamics between at least two actors—if not more. The absence of these interactions in a survey setting limits this method's contextual realism. This limitation is not confined to surveys, of course. As we note above, some cyber wargames also lack strategic interaction. However, for surveys, this limitation is innate to the medium, rather than the result of a design choice.

*Analytical Wargaming*

Wargaming as an analytical method has the potential to balance some of the otherwise stark analytical tradeoffs made by surveys as well as modeling and simulation. First, wargames can offer a reasonable approximation of "the real world." As outlined above, cyber wargames can be calibrated to be as complex or as simple as necessary to engage with their respective research questions. Second, wargames incorporate human actions and responses—decisions—that modeling and simulation cannot. They also bring in strategic play between humans that surveys omit. Relatedly, wargames work to overcome concerns regarding the lack of consequences facing participants when they take part in a survey experiment. Players must deal with the ramifications of their actions. Though as researchers we are rarely concerned with who "wins" or "loses," the competitive aspect of wargames encourages participant buy-in and, as a result, yields behavior more reflective of real-life situations.

In the simplex bounded by analytical utility, contextual realism, and engaged play, we submit that analytical wargames might represent something of a Goldilocks method. Moreover, they may be used to help overcome some of the weaknesses of any given data-generating process as part of a multi-method research design. [Insert: Figure 3 shows each method previously discussed placed on the trilemma's plane. As a whole, analytical wargaming occupies a place in the center. Individual wargames make different tradeoffs, as we argue, shifting their placement in one direction or the other. These tradeoffs also indicate how best to pair individual wargames with other methods. Games designed to hew closely to analytical utility could benefit from pairing with empirical data, for example, while games more focused on engaging play or contextual realism might benefit from a supporting simulation or survey.

*[Insert: Figure 3: Methods and the Trilemma]*

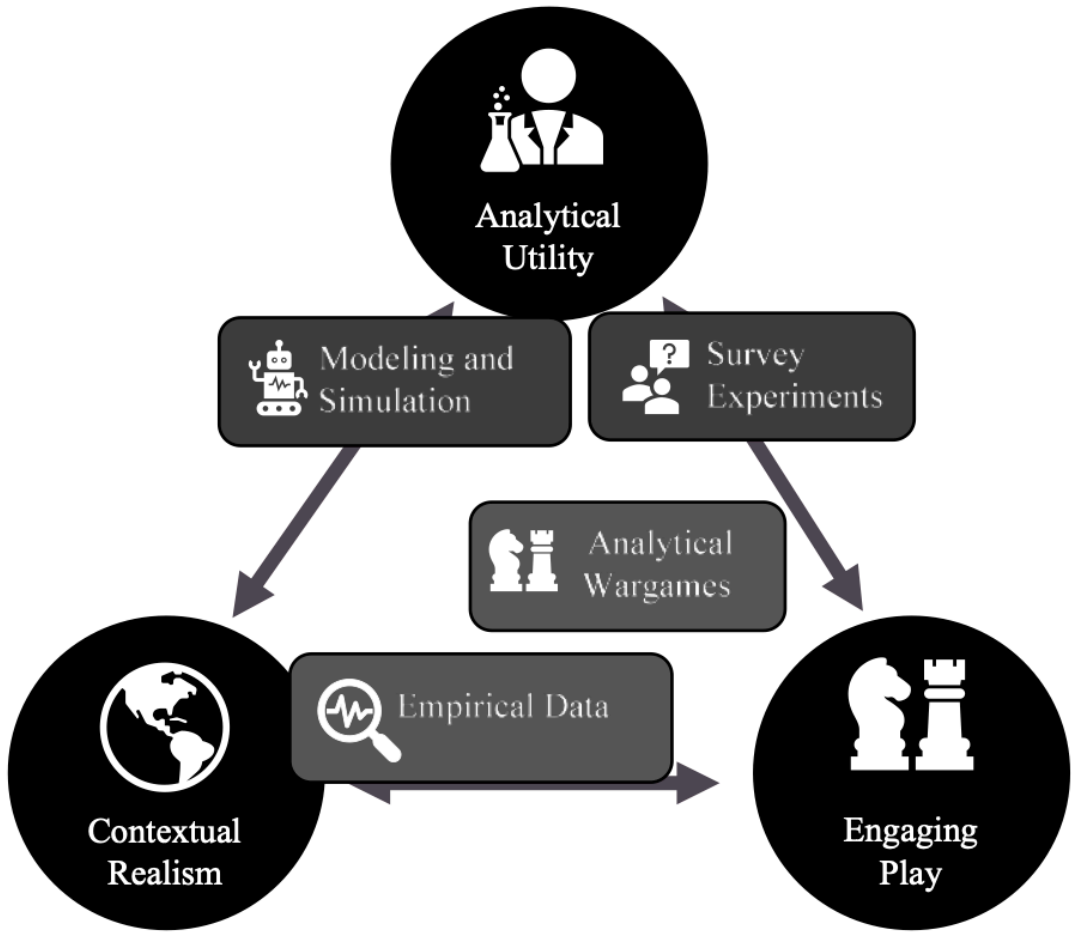

Figure 3: Methods and the Trilemma

**The Promise of Cyber Wargaming**

In this chapter, we outline the wargamer's trilemma and the trade-offs between analytical utility, contextual realism, and engaged play to interrogate both cyber wargames and other methods for generating synthetic data. Our framework reflects the potential of wargaming for cyber research, as well as the need to think critically about the design trade-offs when using wargames as an analytical tool. Methodological work that further interrogates the tradeoffs described by the wargamer's trilemma, in our estimation, has an important role to play advancing this field's ability to address important research questions beyond the reach of other methods.

In our own work, we use wargaming and survey methods to examine deterrence in the cyber domain.[35] Specifically, we examine when and under what conditions players make deterrent threats matched with a particular domain, when these deterrent threats succeed, and whether players attempt to deter cyber-attacks in a meaningfully different manner than attacks emanating from other domains. In the process, we have had to weigh the balance between analytical utility, contextual realism, and engaged play—ultimately creating a game that appropriately models both the threat space and the action space.

By extension, we also engage with a renewed scholarly debate concerning the relative costs and benefits of revealing or concealing clandestine—in this case cyber—capabilities to an adversary.[36] This literature is largely theoretical, not surprisingly, reflecting the paucity of open-source data on clandestine capabilities in general. Our preliminary results do suggest that players treat cyber capabilities differently than conventional and nuclear capabilities—and that the potential to reveal information to an adversary regarding a particular cyber capability may reduce the likelihood of making a threat in the first place.

As more scholars engage with this method of research and analysis, it's our hope that the library of cyber wargames and the new knowledge they can create will continue to grow.

[1] Buchanan, Ben. *The cybersecurity dilemma: Hacking, trust, and fear between nations*. Oxford University Press, 2016; Buchanan, Ben, and Fiona S. Cunningham. "Preparing the Cyber Battlefield: Assessing a Novel Escalation Risk in a Sino-American Crisis (Fall 2020)." *Texas National Security Review* (2020).

[2] Jensen, Benjamin, and Brandon Valeriano. "Cyber Escalation Dynamics: Results from War Game Experiments International Studies Association, Annual Meeting Panel: War Gaming and Simulations in International Conflict March 27, 2019." (2019). Kreps, Sarah, and Jacquelyn Schneider. "Escalation firebreaks in the cyber, conventional, and nuclear domains: Moving beyond effects-based logics." *Journal of Cybersecurity* 5, no. 1 (2019); Borghard, Erica D., and Shawn W. Lonergan. "Cyber operations as imperfect tools of escalation." *Strategic Studies Quarterly* 13, no. 3 (2019): 122-145.

[3] There are several efforts to capture cyber incident data, such as the Dyadic Cyber Incident and Dispute Data, see Valeriano, Brandon, and Ryan C Maness. 2014. "The Dynamics of Cyber Conflict between Rival Antagonists, 2001–11." *Journal of Peace Research* 51 (3): 347–60. "Significant Cyber Incidents | Center for Strategic and International Studies." n.d. Accessed December 16, 2020. https://www.csis.org/programs/strategic-technologies-program/significant-cyber-incidents; "The Cyber Vault Project." 2016. National Security Archive. February 10, 2016. https://nsarchive.gwu.edu/project/cyber-vault-project; "Tracking State-Sponsored Cyberattacks Around the World." n.d. Council on Foreign Relations. Accessed December 16, 2020. These efforts, though subject to selection effects, are essential to understanding state behavior in the cyber domain.

[4] Reddie, Andrew W., Bethany L. Goldblum, Kiran Lakkaraju, Jason Reinhardt, Michael Nacht, and Laura Epifanovskaya. "Next-generation wargames." *Science* 362, no. 6421 (2018): 1362-1364.

[5] As our colleagues at the Naval Postgraduate School note, wargames are also used for education and training purposes. We focus here primarily upon the analytical applications of wargaming methods: Appleget, Jeff, and Robert Burks. *The Craft of Wargaming: A Detailed Planning Guide for Defense Planners and Analysts*. Naval Institute Press, 2020.

[6] Schoenmaker, Dirk. "The financial trilemma." *Economics letters* 111, no. 1 (2011): 57-59.

[7] McGrath, Joseph E. "Dilemmatics: The study of research choices and dilemmas." American Behavioral Scientist 25, no. 2 (1981): 179-210.

[8] Perla, Peter P. *The art of wargaming: A guide for professionals and hobbyists*. Naval Institute Press, 1990. For an alternative interpretation, Andrew W. Reddie et al., "Next-Generation Wargames," *Science* 362, no. 6421 (2018).

[9] Reddie, Andrew W. and Bethany L. Goldblum, "Evidence of the Unthinkable: Experimental Wargaming at the Nuclear Threshold," *Journal of Peace Research*, forthcoming.

[10] Lin-Greenberg, Erik, Reid BC Pauly, and Jacquelyn G. Schneider. "Wargaming for international relations research." *European Journal of International Relations* 28, no. 1 (2022): 83-109.

[11] Silverman, Kaja. *The Subject of Semiotics*. New York: Oxford University Press, 1989.

[12] Salen, K., and E. Zimmerman. "Rules of play-game design fundamentals." MIT Press. (2003); This is referred to as "consequence-based outcomes" in Lin-Greenberg, Erik, Reid Pauly, and Jacquelyn Schneider. 2020. "Wargaming for Political Science Research." SSRN Scholarly Paper ID 3676665. Rochester, NY: Social Science Research Network.

[13] Doing this work, we argue, is central to addressing "the technical and infrastructural perplexities" associated with cyberspace and wargaming: Kollars, Nina and Benjamin Schechter. "Pathologies of Obfuscation: Nobody Understands Cyber Operations or Wargaming," Atlantic Council. February 2021.

[14] Salen, K., and E. Zimmerman. "Rules of play-game design fundamentals." MIT Press. (2003)

[15] Schechter, Benjamin, Jacquelyn Schneider, and Rachael Shaffer. "Wargaming as a Methodology: The International Crisis Wargame and Experimental Wargaming." *Simulation & Gaming*, (January 2021).

[16] Reddie, Andrew W., Bethany L. Goldblum, Kiran Lakkaraju, Jason Reinhardt, Michael Nacht, and Laura Epifanovskaya. "Next-generation wargames." *Science* 362, no. 6421 (2018): 1362-1364; Lin-Greenberg, Erik. "Wargame of Drones: Remotely Piloted Aircraft and Crisis Escalation." *Available at SSRN 3288988* (2020).

[17] Jensen, Ben and David Banks, "Cyber Operations in Conflict: Lessons from Analytic Wargames - CLTC UC Berkeley Center for Long-Term Cybersecurity." n.d. CLTC (blog). Accessed January 27, 2021. https://cltc.berkeley.edu/cyber-operations/.

[18] This survey experiment also draws on conclusions drawn from *Netwar*, a game designed to collect the dynamics between a government, a violent non-state actor, a major international firm, and a cyber activist network in the cyber domain.

[19] Jensen, Benjamin, and Brandon Valeriano. "Cyber Escalation Dynamics: Results from War Game Experiments," International Studies Association, Annual Meeting Panel: War Gaming and Simulations in International Conflict March 27, 2019.

[20] Jensen, Benjamin, and Brandon Valeriano. "Cyber Escalation Dynamics: Results from War Game Experiments," International Studies Association, Annual Meeting Panel: War Gaming and Simulations in International Conflict March 27, 2019. p. 11.

[21] Reddie, Andrew W., Bethany L. Goldblum, Kiran Lakkaraju, Jason Reinhardt, Michael Nacht, and Laura Epifanovskaya. "Next-generation wargames." *Science* 362, no. 6421 (2018): 1362-1364.

[22] Kello, Lucas. "The meaning of the cyber revolution: Perils to theory and statecraft." *International Security* 38, no. 2 (2013): 7-40; Buchanan, Ben, and Fiona S. Cunningham. "Preparing the Cyber Battlefield: Assessing a Novel Escalation Risk in a Sino-American Crisis (Fall 2020)." *Texas National Security Review* (2020); Buchanan, Ben. *The cybersecurity dilemma: Hacking, trust, and fear between nations*. Oxford University Press, 2016; Borghard, E. D., & Lonergan, S. W. (2019). Cyber operations as imperfect tools of escalation. *Strategic Studies Quarterly*, *13*(3), 122-145.

[23] Acton, James M. "Cyber warfare & inadvertent escalation." *Dædalus* 149, no. 2 (2020): 133-149; Buchanan, Ben, and Fiona S. Cunningham. "Preparing the Cyber Battlefield: Assessing a Novel Escalation Risk in a Sino-American Crisis." *Intelligence* 3, no. 4 (2020): 54-81; Harknett, Richard J., and Max Smeets. "Cyber campaigns and strategic outcomes." *Journal of Strategic Studies* (2020): 1-34.

[24] For instance, Bao, Tiffany, Yan Shoshitaishvili, Ruoyu Wang, Christopher Kruegel, Giovanni Vigna, and David Brumley. 2017. "How Shall We Play a Game?: A Game-Theoretical Model for Cyber-Warfare Games." In *2017 IEEE 30th Computer Security Foundations Symposium (CSF)*, 7–21. Or, for a survey of existing methods, see: Wang, Yuan, Yongjun Wang, Jing Liu, Zhijian Huang, and Peidai Xie. 2016. "A Survey of Game Theoretic Methods for Cyber

Security." In *2016 IEEE First International Conference on Data Science in Cyberspace (DSC)*, 631–36.

[25] Kotenko, Igor, and Andrey Chechulin. "A cyber attack modeling and impact assessment framework." In *2013 5th International Conference on Cyber Conflict (CYCON 2013)*, pp. 1-24. IEEE, 2013.; Ben-Asher, Noam, and Cleotilde Gonzalez. "CyberWar game: a paradigm for understanding new challenges of cyber war." In *Cyber Warfare*, pp. 207-220. Springer, Cham, 2015.

[26] Marchi, Scott de, and Scott E. Page. 2014. "Agent-Based Models." *Annual Review of Political Science* 17 (1): 1–20.

[27] Put in blunt terms, analysts should be wary of the "garbage in, garbage out" problem associated with any synthetic data generating process.

[28] Veksler, Vladislav D., Norbou Buchler, Claire G. LaFleur, Michael S. Yu, Christian Lebiere, and Cleotilde Gonzalez. 2020. "Cognitive Models in Cybersecurity: Learning From Expert Analysts and Predicting Attacker Behavior." *Frontiers in Psychology* 11.

[29] Shandler, Ryan, Michael L. Gross, and Daphna Canetti. "A fragile public preference for cyber strikes: Evidence from survey experiments in the United States, United Kingdom, and Israel." *Contemporary Security Policy* (2021): 1-28; Gross, Michael L., Daphna Canetti, and Dana R. Vashdi. "The psychological effects of cyber terrorism." *Bulletin of the Atomic Scientists* 72, no. 5 (2016): 284-291.

[30] Kreps, Sarah, and Jacquelyn Schneider. "Escalation firebreaks in the cyber, conventional, and nuclear domains: Moving beyond effects-based logics." *Journal of Cybersecurity* 5, no. 1 (2019): tyz007.

[31] Kreps, Sarah E., and Debak Das. "Warring from the Virtual to the Real: Assessing the Public's Threshold for War on Cyber Security." *Available at SSRN 2899423* (2017); Kostyuk, Nadiya, and Carly Wayne. "The microfoundations of state cybersecurity: Cyber risk perceptions and the mass public." *Journal of Global Security Studies* (2020).

[32] Oberholtzer, Jenny, Abby Doll, David Frelinger, Karl Mueller, and Stacie Pettyjohn. "Applying wargames to real-world policies." *Science* 363, no. 6434 (2019): 1406-1406.

[33] Kertzer, Joshua D. "Re-Assessing Elite-Public Gaps in Political Behavior." *American Journal of Political Science* (2020).

[34] Fowler Jr, Floyd J., and Floyd J. Fowler. Improving survey questions: Design and evaluation. Sage, 1995.

[35] Libicki, Martin C. "Expectations of cyber deterrence." Strategic Studies Quarterly 12, no. 4 (2018): 44-57; Goodman, Will. "Cyber deterrence: Tougher in theory than in practice?." Strategic Studies Quarterly 4, no. 3 (2010): 102-135; Tor, Uri. "'Cumulative deterrence' as a new paradigm for cyber deterrence." Journal of Strategic Studies 40, no. 1-2 (2017): 92-117; Wilner, Alex S. "US cyber deterrence: Practice guiding theory." Journal of Strategic Studies 43, no. 2 (2020): 245-280; Brantly, Aaron F. "The cyber deterrence problem." In 2018 10th International Conference on Cyber Conflict (CyCon), pp. 31-54. IEEE, 2018; Crosston, M. D. (2011). World gone cyber MAD: How "mutually assured debilitation" is the best hope for cyber deterrence. Strategic studies quarterly, 5(1), 100-116; Lupovici, Amir. "The "Attribution Problem" and the social construction of "Violence": Taking cyber deterrence literature a step forward." International Studies Perspectives 17, no. 3 (2016): 322-342; Klimburg, Alexander. "Mixed Signals: A Flawed Approach to Cyber Deterrence." Survival 62, no. 1 (2020): 107-130.

[36] Green, Brendan Rittenhouse, and Austin Long. "Conceal or Reveal? Managing Clandestine Military Capabilities in Peacetime Competition." *International Security* 44, no. 3 (2020): 48-83.